\documentclass[english,5p,twocolumn,times]{elsarticle}
\usepackage[T1]{fontenc}
\usepackage[utf8]{inputenc}
\usepackage{geometry}
\usepackage{babel}
\usepackage{verbatim}
\usepackage{graphicx}
\usepackage[unicode=true]
 {hyperref}

\makeatletter
\usepackage{color}
\usepackage{graphics}
\usepackage{array}
\usepackage{graphicx}
\usepackage{lastpage}
\usepackage{float}

\usepackage{times}

\usepackage{lineno}

\journal{Artificial Intelligence in Medicine}

\usepackage{amsmath,amssymb}

\makeatother

\begin{document}
\sloppy
\begin{frontmatter}

\title{A data science approach to drug safety:\\
Semantic and visual mining of adverse drug events from clinical trials
of pain treatments}

\author[label1,label2]{Jean-Baptiste Lamy\corref{cor1}}
\ead{jibalamy@free.fr}
\cortext[cor1]{Corresponding author\\This is an author file of the article published in Artificial Intelligence In Medicine 2021;115:102074, DOI: 10.1016/j.artmed.2021.102074 ; it is available under Creative Commons Attribution Non-Commercial No Derivatives License. This is also an updated version of Arxiv preprint arxiv:2006.16910 }

\address[label1]{Université Sorbonne Paris Nord, LIMICS, Sorbonne Université, INSERM, UMR 1142, F-93000, Bobigny, France}
\address[label2]{Laboratoire de Recherche en Informatique, CNRS/Université Paris-Sud/Université Paris-Saclay, Orsay, France}
\begin{abstract}
Clinical trials are the basis of Evidence-Based Medicine. Trial results
are reviewed by experts and consensus panels for producing meta-analyses
and clinical practice guidelines. However, reviewing these results
is a long and tedious task, hence the meta-analyses and guidelines
are not updated each time a new trial is published. Moreover, the
independence of experts may be difficult to appraise. On the contrary,
in many other domains, including medical risk analysis, the advent
of data science, big data and visual analytics allowed moving from
\emph{expert}-based to \emph{fact}-based knowledge. Since 12 years,
many trial results are publicly available online in trial registries.
Nevertheless, data science methods have not yet been applied widely
to trial data.

In this paper, we present a platform for analyzing the safety events
reported during clinical trials and published in trial registries.
This platform is based on an ontological model including 582 trials
on pain treatments, and uses semantic web technologies for querying
this dataset at various levels of granularity. It also relies on a
26-dimensional flower glyph for the visualization of the Adverse Drug
Events (ADE) rates in 13 categories and 2 levels of seriousness. We
illustrate the interest of this platform through several use cases
and we were able to find back conclusions that were initially found
during meta-analyses. The platform was presented to four experts in
drug safety, and is publicly available online, with the ontology of
pain treatment ADE.
\end{abstract}
\end{frontmatter}

\section{Introduction}

Clinical trials are the basis of Evidence-Based Medicine (EBM) \citep{Sheridan2016}.
In particular, they provide evidence of the efficacy and the safety
of drug treatments. Trial results are reviewed by medical experts
and consensus panels during the process of performing meta-analyses
and writing clinical practice guidelines. These processes remain widely
manual and based on human expertise.

However, reviewing trial results is a tedious task, and the independence
of experts is somehow questionable \citep{Moynihan2008}, \emph{e.g.}
it has been shown that up to 90\% of guidelines authors have ties
to drug firms \citep{Cosgrove2013}. Independent experts are rare:
to be an expert, one has to work on industry-funded trials, and disclosing
links of interests does not necessarily prevent biases \citep{Elliott2009}.
Beyond independence, human expertise is not reproducible \citep{Baker2016},
leading to variability in the recommendations across countries and
organizations \citep{Cosgrove2013}. Finally, the analysis of trial
data relies heavily on statistical methods that have known limits
\citep{Gelman2006,Cerrito2011}, \emph{e.g.} a significant difference
may actually be very small.

For example, a recent meta-analysis on tapentadol, a new opioid drug
for acute pain, included 8 randomized clinical trials and 3,706 patients,
and showed that tapentadol was associated with fewer gastrointestinal
adverse drug events (ADE) \citep{Wang2020}. However, such a meta-analysis
requires months of tedious expert work \citep{Valkenhoef2012}, and
is not updated each time a new trial is published. Moreover, clinicians
might have difficulties to assess the quality of the study and the
independence of the authors, despite two investigators were involved
in the process. Indeed, lack of confidence and trust, and lack of
time to appraise evidence have been identified as barriers to the
use of EBM by GPs \citep{Zwolsman2012}.%

In many domains, the advent of data science, big data and visual analytics
allowed moving from \emph{expert}-based to \emph{fact}-based knowledge
\citep{Daniel2016}. These methods have been shown to be efficient
for the analysis of medical risk \citep{Cerrito2011}. Today, regarding
clinical trials, a lot of data is publicly available. Trial registration
is mandatory since 2005 (International Committee of Medical Journal
Editors) and 2008 (revised Declaration of Helsinki). Moreover, the
publication of most trial results is mandatory in the US since 2017
(FDAAA 801, Final Rule). In May 2020, more than 42,000 study results
are available in ClinicalTrials.gov, the largest trial registry.

Nevertheless, methods from the field of data science have not yet
been widely applied to clinical trial data. This may be explained
by the fact that the full, per-patient, outcomes data are not publicly
available. Only aggregated data are available in trial registries,
such as the number of patients having a given ADE or the mean value
of a biological marker. However, data science methods are often not
suited for such aggregated data, and in particular the data aggregation
prevents the application of machine learning algorithms, \emph{e.g.}
one cannot learn a model to predict patient outcomes from patient
characteristics using aggregated data. Consequently, the automated
computer-based treatment of publicly available trial data is particularly
challenging.

In previous studies, we showed that semantic web technologies and
visual analytics were an interesting option for accessing and comparing
drug knowledge \citep{Lamy2017} and for ranking and visualizing the
properties of antibiotic agents \citep{Tsopra2018_2,Lamy2020_2,Tsopra2019_2}.
Here, our hypothesis is that similar methods could be applied to aggregated
trial result data found in trial registries, and could complement
the currently used statistical methods.

Information visualization and visual analytics have been proposed
as a solution to deal with mass of data in medicine \citep{visumed-infovisumed}.
Two strategies can be distinguished when designing visual analytics-based
systems. The first one consists of using well-known generic visual
techniques, often in combination. For example, W. Wang \emph{et al.}
\citep{Wang2020_2} applied interactive bubble charts and scatter
plots to non-aggregated trial safety data. H. Ltifi \emph{et al.}
\citep{Ltifi2020} combined graphs and 3D bar charts for the analysis
of nosocomial infections in intensive care units. D.J. Feller \emph{et
al.} \citep{Feller2018} combined heatmaps and density plots in Glucolyzer,
a tool for helping dieticians identify patterns between blood glucose
levels and meal composition in type 2 diabetes. ClinOmicsTrailbc \citep{Schneider2019}
combined radar charts, scatter plots, histograms and circular sun
plots for breast cancer treatment stratification. The second strategy,
more rarely employed, consists of designing visual approaches that
are specifically adapted to the desired domain. As stated by L. Chittaro
\citep{visumed-infovisumed}, a key research problem is to discover
new visual metaphors for representing medical information and to understand
what task they can support. For example, CareVis \citep{visumed-gbp-plan}
proposed a novel PlanningLine glyph for visualizing temporal care
plan. J.M. Juarez \emph{et al.} \citep{Juarez2015} designed a multiple
temporal axis model for the visualization of the activity of a single
patient for homecare monitoring. J. Bernard \emph{et al.} \citep{Bernard2019}
designed a specific dashboard for representing patient history, and
used dashboard networks for visualizing multiple patients histories.
Here, we will propose a specific visual metaphor for viewing and comparing
ADE rates, grouped in 13 categories.

In the previous example on tapentadol, data science methods may be
used to easily and automatically produce rapid results on ADE rates.
A data mining platform may contain all ADE observed during trials
with publicly available results. A semantic search engine may allow
to automatically search for trials testing tapentadol or other opioids
for acute pain. Finally, the platform may aggregate and compare the
results of the retrieved trials, using specifically designed visual
analytics, showing within minutes that tapentadol has a lower risk
of digestive ADE. Of course, the generated evidence will still need
to be interpreted, and it would not have the strength of a full meta-analysis.
Nevertheless, such a platform could help experts while performing
meta-analyses, allowing to quickly test various hypotheses (\emph{e.g.}
what about comparing tapentadol to morphine?). In addition, it could
also be used by non-expert clinicians for verifying the results of
a published meta-analysis and assessing its reproducibility, or for
obtaining up-to-date results including the latest studies.

The objective of this work is to design a web platform for the semantic
and visual mining of ADE observed in clinical trials and published
in trial registries, and to apply this platform to pain treatments.
This platform aims at helping experts and clinicians, but also at
illustrating what data science applied to trial public data may bring
to drug safety.

The rest of the paper is organized as follows. Section \ref{sec:Related-works}
describes related works on trial data and visualization. Section \ref{sec:Materials_and_methods}
describes the methods used for building the platform, including trial
selection, ontology modeling and population, ontology querying, data
correction and normalization, and visualization. Section \ref{sec:Results}
presents the resulting ontology of pain treatment ADE and the proposed
platform, and details several use cases and the comments of experts
in drug safety. Section \ref{sec:Discussion} discusses the methods
and the results, and, finally, section \ref{sec:Conclusion} concludes
with perspectives.

\section{\label{sec:Related-works}Related works}

\subsection{Usage of clinical trial registries}

Clinical trial data is available publicly in online registries, such
as ClinicalTrials.gov \citep{Zarin2011}. For some trials, it includes
trial results, with the list of ADE observed in the various patient
groups of the study. Today, trial registries are used for systematic
reviews \citep{Honig2010} and network meta-analyses \citep{Zarin2017,Rudroju2013},
comparing several treatments by chaining trial results. But the efforts
to standardize information from trials have not yet resulted in improvements
in the dissemination of trial evidence \citep{Valkenhoef2012}.

Similarly, in drug safety, many works focused on clinical data collection
from EHR and hospitals, but a recent review \citep{Natsiavas2019}
highlighted that most of these works focused on extracting, representing
and integrating information, rather than the use and the dissemination
of this information. For instance, C. Zhan \emph{et al.} applied computational
methods to prescription data for detecting ADE signals \citep{Zhan2020}.
On the contrary, clinical trial data is mostly analyzed with statistical
methods, and few computational methods were proposed.

A few pioneering works focused on the direct use of trial data. I.
Atal \emph{et al.} proposed a visual approach for viewing on a mapping
the research effort and the health needs of low-income regions \citep{Atal2018}.
J. Warner \emph{et al.} proposed a network visualization of chemotherapy
treatment regimens \citep{Warner2013}. Z. He \emph{et al.} proposed
the use of text mining, bar charts and diagrams for the analysis of
clinical trial target populations \citep{He2015}. Finally, J. Sjöbergh
\emph{et al.} combined maps, parallel coordinates and diagrams for
visualizing the individual patient data in a single trial \citep{Sjobergh2012}.

Ontologies and semantic web technologies consist of formal and unambiguous
models; they have been widely applied to the medical domain \citep{Schulz2013}.
An ontology was proposed for structuring clinical trial data, named
OCRe \citep{Sim2013}. However, this ontology is very complex, and
no tool exists for the automatic population of the ontology from trial
registries.

\subsection{\label{subsec:Multidimensional-visualization-t}Multidimensional
visualization techniques}

Many techniques exist for the visualization of multidimensional numeric
data; in this section we will briefly review the three main approaches.
First, \emph{dimension reduction} consists of reducing the number
of dimensions to 2 or 3, at the price of an information loss. Then,
the data can be visualized using a simple scatter plot. The main dimension
reduction techniques are Principal Component Analysis (PCA), Multidimensional
Scaling (MDS) \citep{Borg2013_2}, Self-Organizing Map (SOM) \citep{Kohonen1995}
and t-distributed Stochastic Neighbor Embedding (t-SNE) \citep{van2008}.
Dimension reduction techniques are particularly useful for grouping
similar items in clusters. t-SNE is commonly used in bioinformatics,
\emph{e.g.} for the visualization of transcriptomic \citep{Cieslak2020},
but in medicine, it was also used for the classification of patients
with Parkinson's disease \citep{Oliveira2018}.

Second, parallel coordinates \citep{Inselberg2009} consist of representing
each dimension by a parallel axis (usually vertical), and each data
point by a broken line that crosses each axis at the corresponding
value of the point in that dimension. Parallel coordinates are often
associated with interactive interfaces allowing the selection of a
subset of data points. They are particularly good at facilitating
the discovery of patterns across the dimensions. Recently, parallel
coordinates were proposed for the study of exposure to oxides of nitrogen
and its relation to adverse birth outcomes \citep{Mitku2020}. Other
applications include the visualization of multi-omics networks \citep{Kanai2018}.

\begin{figure}
\noindent \begin{centering}
\includegraphics[width=1\columnwidth]{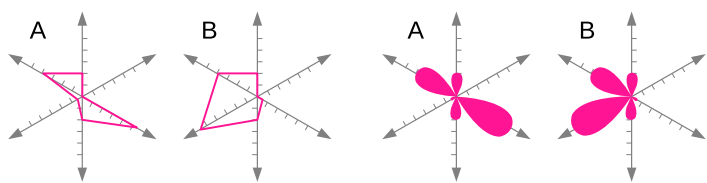}
\par\end{centering}
\caption{\label{fig:ex_glyphs}Examples of star glyphs (left) and flower glyphs
(right). Dimension axes are commonly omitted, especially in flower
glyphs.}
\end{figure}

Third, glyphs consist of representing each data point by a small icon
called a glyph, and each dimension by a characteristic of the glyph,
\emph{e.g.} the length or the color of a given element \citep{Ward}.
Two kinds of glyphs can be distinguished: metaphoric glyphs that aim
at looking like a common object, often related to the data visualized
but not necessarily, and abstract glyphs that are geometric construction
with no similarity to any common object. An example of metaphoric
glyph is Chernoff's faces \citep{visu-face-chernoff}: these glyphs
look like human faces. An example of abstract glyph is the ``stardinates''
\citep{visu-glyphe-stardinate} or star glyphs \citep{Fuchs2014}
(see examples in Figure \ref{fig:ex_glyphs}), which are similar to
parallel coordinates but with the coordinates organized in a star.
Each dimension is represented by an axis that crosses at the center,
and each data point is represented by a polygon. Contrary to parallel
coordinates, each data point is represented in a separate glyph. Flower
glyphs \citep{Chau2011,Pilato2016,Keck2017,Keck2018} are close to
star glyphs but follows a ``flower'' metaphor : each glyph is a
flower having one petal per dimension, the size of the petal being
proportional to the value of the data point in that dimension. An
interest of flower glyphs is that they are less dependent on the order
in which the dimensions are displayed. In Figure \ref{fig:ex_glyphs},
glyphs A and B are the same but the order of the dimensions differs.
Notice that A and B have the same inner area with flower glyphs, but
not with star glyphs.

Glyphs have shown their ability to visualize large datasets with hundreds
of dimensions \citep{visumed-glyph-100dimensions}. In medicine, they
have been used for facilitating the analysis of semen \citep{Duffy2015},
for example. Glyphs are of particular interest for identifying similarity
or differences between data point.

\section{\label{sec:Materials_and_methods}Materials and methods}

\subsection{\label{sec:Materials}Materials}

Clinical trials were searched and extracted from ClinicalTrials.gov.
We focused on pain, but we also considered fever, because of the large
overlap between the two, both in terms of symptoms (fever is often
associated with pain) and drugs (\emph{e.g.} paracetamol and ibuprofen
are both antipyretics and painkillers). We used the following query:
Condition (\emph{i.e.} ``pain'' or ``fever'') + Completed Studies
+ Studies With Results + Phase 3 or 4. Searches were performed on
the 18\textsuperscript{th} January 2020.

From the resulting list of trials, we manually excluded: (1) trials
not involving the treatment of pain or fever (\emph{e.g.} trial evaluating
the pain associated with the injection of a given vaccine, in which
pain is an outcome measure but not an indication, or trials focused
on insomnia caused by chronic low back pain), (2) trials whose protocol
does not allow comparing treatments (\emph{e.g.} trials comparing
two care protocols where each patient in a group did not receive the
same treatment, or trials during which drug labeling errors occurred),
(3) trials testing non-drug treatment (\emph{e.g.} behavioral training
or surgery), (4) trials testing homeopathic drugs, (5) trials where
pain is treated by disorder-specific drugs (\emph{e.g.} chest pain
caused by angina pectoris and treated by cardiac drugs), and (6) trials
testing anesthetic agents or painkillers used during surgical operation.
Trials comparing a drug treatment with a non-drug treatment were included,
but only the groups receiving drug treatments were considered in the
present study.

\subsection{Ontology modeling}

We designed an ontology of pain treatment ADE in clinical trials.
Its purpose was not to fully model the domain of clinical trials,
as does OCRe for example. On the contrary, it aimed at being a simple
model, limited to the needs of data mining of ADE, and allowing us
to handle ADE data as semantically linked data. The use of the ontological
formalism was motivated by the reuse of existing tools designed for
ontologies, and by the fact that ontologies are good at dealing with
inheritance, \emph{i.e.} various levels of granularity. Here, we widely
used inheritance in the modeling of indications (\emph{e.g.} post-vaccination
fever is more specific than fever) and active principles (\emph{e.g.}
ibuprofen is more specific than NSAI, Non-Steroid Anti-Inflammatory
drugs). This will allow querying the ontology at various levels of
granularity.

In the ontology, the central class, Group, represents a group of similar
patients, receiving the same treatment(s) for the same indication.
ADE are observed in groups. On the contrary, individual patients are
not present in the ontology. ADE were classified in 2 levels of seriousness
and 13 categories, including 12 anatomo-functional categories and
1 ``Unclassified'' category for general symptoms (such as fatigue
or unspecified infectious diseases). These categories are more general
than the 27 MedDRA top-level System Organ Classes (SOC). A given ADE
was allowed to belong to more than one category, \emph{e.g.} allergic
rhinitis belongs to both respiratory system and blood and immune system.

We manually mapped the MedDRA terms to these 13 categories. Most MedDRA
SOC could directly be mapped to a single category (\emph{e.g.} respiratory
system). However, the ``Investigations'' SOC required more work,
in order to associate each abnormal test result to the right category
(\emph{e.g.} ``Tidal volume decreased'' was associated with the
respiratory system category).

The resulting ontology was interfaced with Python scripts using the
Owlready 2 ontology-oriented programming module \citep{Lamy2017_5,Lamy2016_3}.
The ontology was stored in the Owlready 2 quadstore, as an SQLite3
database.

\subsection{Ontology population}

The pain treatment ADE ontology was populated from ClinicalTrials.org
XML data, using a semi-automatic process. In ClinicalTrials.org, trials,
groups and ADE observations are well structured, and ADE are classified
according to the 27 System Organ Class (SOC) of the MedDRA terminology,
and in 2 classes of seriousness (serious \emph{vs} non-serious). In
most cases, the ADE label corresponds to a MedDRA term. Consequently,
we automatically extracted and inserted these pieces of data in the
ontology, using Python scripts.

On the contrary, the treatments received by the patient groups are
not coded in ClinicalTrials.org, but only present in free-text fields.
Similarly, the precise indications of these treatments are available
only in free text; although MeSH (Medical Subject Headings) terms
are provided, they often remain too general (\emph{e.g.} pain without
more precision). Thus, treatments and indications cannot be extracted
without a manual intervention.

We wrote Python scripts for extracting the free-text values, and automatically
recognizing named entities and dose regimens. The output was formatted
in CSV files (Comma-Separated Values). Then these files were opened
in a spreadsheet software (LibreOffice Calc) and manually reviewed
by a pharmacist working in medical informatics (JBL). The entire process
took about 1 month.

A first script aimed at automatically detecting the names of active
principles (using a list extracted from UMLS, Unified Medical Language
System), their dose, dose unit and number of intakes per day (using
regular expressions targeting common expressions such as ``bid''
: \emph{bis in die}, \emph{i.e.} twice per day). A second script aimed
at automatically extracting indication, from trial summary, description
and MeSH terms. We measured the performance of the two scripts, by
comparing the data extracted automatically by the scripts with the
same data after correction during the manual review.

When coding doses and numbers of intakes per day, we allowed the use
of a range with a minimum and a maximum value (\emph{e.g.} 5-10 mg
or 1-2 intakes per day). In addition, when the dose varied over time,
we kept only the maintenance dose.%

The active principles and indications identified were added to the
ontology, and hierarchically structured. Active principles were classified
according to three dimensions: chemical structure (\emph{e.g.} steroid),
pharmacological activity (\emph{e.g.} antihistamine) and main indication
(\emph{e.g.} anti-epileptics). Indications were classified according
to four dimensions: anatomy (\emph{e.g.} musculoskeletal pain), etiology
(\emph{e.g.} neuropathic pain), chronicity (\emph{i.e.} chronic or
acute pain) and severity (\emph{i.e.} mild, moderate or severe pain).
These classifications were highly inspired by existing terminologies
such as ATC (Anatomical Therapeutical Chemical classification of drugs)
or ICD10 (International Classification of Diseases, release 10).

ADE terms were extracted from ClinicalTrials.gov and automatically
mapped to MedDRA terms using their textual English label. When no
corresponding MedDRA term was found, the term was simply associated
with the MedDRA SOC present in ClinicalTrials.gov. While more general,
the SOC still permits relating the ADE to one of our 13 categories
of ADE.

Several trials include a titration period, mainly for opioid painkillers.
A typical design study for comparing a given opioid with a placebo
is as follows: an open-label titration period with the test drug,
including all patients, then randomization followed by a maintenance
period with two groups, one taking the test drug and the other taking
a placebo. In this case, we chose to include in the ontology the ADE
observed during the titration period, but without mixing them with
those observed in the maintenance period, because there is no titration
period for placebo and thus no comparison is possible.

The ontology population was performed by a pharmacist working in medical
informatics (JBL), and took about 1 month.

\subsection{Ontology querying}

We designed a query procedure for searching the ontology. The procedure
takes as input one or more group definitions. Each group definition
may include criteria regarding the trial (\emph{e.g.} restrict to
randomized trials), the indication (\emph{e.g.} restrict to cancer
pain) and the active principles (\emph{e.g.} opioid or morphine).
The ontology allows the use of various levels of granularity in the
formulation of the query, as in the latter example. Several active
principles may be mentioned for a given group (corresponding to a
bi- or tritherapy), and, for each, a specific release (immediate or
modified), range of dose (\emph{e.g.} 5-10 mg) and number of intakes
per day (\emph{e.g.} 1-2 times per day) may be specified.

In addition, we considered two particular situations when querying
on active principles. First, we also allowed an ``open list'' search,
that returns groups with the active principles queried possibly in
association with others, \emph{e.g.} morphine alone or associated
with any other active principle. This ``open list'' search comes
in addition to the usual ``closed list'' search, which is the default.
Second, it is sometimes interesting to perform a comparison between
two treatments defined at different levels of granularity, \emph{e.g.}
to compare tapentadol \emph{vs} opioids. In this comparison, ``opioids''
implicitly means ``opioids other than tapentadol'', since tapentadol
is an opioid. Thus, we also supported exclusion in the search process.

For a single query, the procedure returns three sets of results. The
first one, \emph{direct comparisons}, includes only direct comparisons,
\emph{i.e.} trials in which all the groups queried are present. The
second one, \emph{direct and indirect comparisons}, includes both
direct comparisons and indirect comparisons \emph{via} placebo, \emph{i.e.}
trials in which only some of the groups queried are present, and where
a placebo group is present for performing an indirect comparison normalized
by placebo (see next section). The third one, \emph{absolute values},
includes all trials containing at least one of the groups queried,
without any correction or normalization.

As a consequence, the first result set (direct comparisons) has the
fewest number of trials and patients but the highest data quality.
On the contrary, the third result set (absolute values) has the highest
number of trials and patients but the lowest data quality. The user
may choose the desired result sets, \emph{e.g.} one may use the direct
comparisons and defaults to the other result sets when there are not
enough patients.

\subsection{\label{subsec:Data-correction}Data correction and normalization}

We implemented three data correction and normalization methods. First,
\emph{per-trial number of patients correction} was implemented for
direct comparisons when more than one trial is involved in the comparison.
For example, let us consider two trials $T_{1}$ and $T_{2}$, both
comparing two drugs $D_{1}$ and $D_{2}$. $T_{1}$ includes 100 patients
treated with $D_{1}$ and 100 patients treated with $D_{2}$, and
$T_{2}$ includes 100 patients treated with $D_{1}$ but 200 patients
treated with $D_{2}$. Without correction, a higher weight is given
to the group $T_{2}$-$D_{2}$, since patients are more numerous in
this group. This is a potential bias if $T_{2}$ is at higher (or
lower) risk of ADE, \emph{e.g.} because the trial involves post-vaccination
fever and the vaccine may cause additional ADE. Consequently, we need
to normalize the data.

We propose to reduce the weight given to each group using a correction
factor, so as the weight given to each group is equivalent to the
weight of the smallest group in the same trial. In the previous example,
the correction factors will be 1.0 for groups $T_{1}$-$D_{1}$, $T_{1}$-$D_{2}$
and $T_{2}$-$D_{1}$, but 0.5 for $T_{2}$-$D_{2}$. This gives an
equivalent weight to $T_{2}$ for both $D_{1}$ and $D_{2}$. In the
general case, for a trial $T$ with $n$ groups $D_{1}$ to $D_{n}$,
the correction factor for group $D_{x}$ is:
\[
w_{x}=\frac{\,\,\min(\left|TD_{i}\right|\,\,for\,\,i\in[1,n])\,\,}{\left|TD_{x}\right|}\,
\]
where $\left|TD_{i}\right|$ is the number of patients in the group
with drug $D_{i}$ in trial $T$.

When computing the correction factor $w_{x}$, we used the minimum
group size and not the average, because using the average would reduce
the weight of the larger groups but also increase the weight of the
smaller groups. In case of very small groups, using the average would
give a disproportional importance to the rare events occurring in
these small groups (\emph{e.g.} 1 stroke in a group of 10 patients).
On the contrary, using the minimum ensures that no patient counts
for ``more than one''.

Second, \emph{placebo normalization} was implemented for indirect
comparisons. In indirect comparisons, the number of ADE observed needs
to be adjusted according to the number of ADE observed in placebo
groups. For example, let us consider an indirect comparison between
two drugs $D_{1}$ and $D_{2}$, using two trials $T_{1}$ and $T_{2}$.
In $T_{1}$, $D_{1}$ is compared to placebo $P$, and, in $T_{2}$,
$D_{2}$ is compared to $P$, each group including 100 patients. Let
us consider $E$, a given ADE, \emph{e.g.} vomiting. We denote by
$E(T_{i}D)$ the number of occurrences of $E$ in trial $T_{i}$ in
the group taking drug $D$. Let $E(T_{1}D_{1})=20$, $E(T_{1}P)=10$,
$E(T_{2}D_{2})=30$, $E(T_{2}P)=30$. These numbers suggest that,
despite more vomitings were observed with $D_{2}$ than $D_{1}$,
$D_{2}$ is at lower risk of causing vomiting because it caused as
much vomitings as placebo, while $D_{1}$ caused more. In facts, the
difference observed between $D_{1}$ and $D_{2}$ is partially due
to the difference in clinical conditions between $T_{1}$ and $T_{2}$.

Here, the average rate of $E$ in the placebo group is $\frac{10+30}{100+100}=20\%$.
But, in $T_{1}$, the rate of $E$ in the placebo group is $\frac{10}{100}=10\%$,
thus there is another $10\%$ missing. Adding these $10\%$ to the
rate of $E$ in the $D_{1}$ leads to a corrected number of occurrences
of $20+0.1\times100=30$. Similarly, for $D_{2}$ in trial $T_{2}$,
the corrected number of occurrences will be $30-0.1\times100=20$.

In the general case, the corrected number of occurrences of $E$ for
the group $T_{x}D_{y}$ is:
\[
E_{c}(T_{x}D_{y})=E(T_{x}D_{y})+\left(\frac{\sum_{i=1}^{n}E(T_{i}P)}{\sum_{i=1}^{n}\left|T_{i}P\right|}-\frac{E(T_{x}P)}{\left|T_{x}P\right|}\right)\times\left|T_{x}D_{y}\right|
\]
where $n$ is the number of trials, $\frac{\sum_{i=1}^{n}E(T_{i}P)}{\sum_{i=1}^{n}\left|T_{i}P\right|}$
is the average rate of $E$ in placebo over all trials, and $\frac{E(T_{x}P)}{\left|T_{x}P\right|}$
the average rate of $E$ in placebo in the considered trial $T_{x}$.

Third, when direct and indirect comparisons are mixed, we need to
ensure that the proportion of patients coming from direct comparisons
is the same in each of the compared groups. Let us consider a mixed
comparison between two drugs $D_{1}$ and $D_{2}$, including $T_{1}$
(direct comparison) with $E(T_{1}D_{1})=40$, $\left|T_{1}D_{1}\right|=100$,
$E(T_{1}D_{2})=50$, $\left|T_{1}D_{2}\right|=100$, but also two
indirect comparisons: $T_{2}$ with $E(T_{2}D_{1})=10$, $\left|T_{2}D_{1}\right|=100$
and $T_{3}$ with $E(T_{3}D_{2})=22$, $\left|T_{3}D_{1}\right|=200$
(after applying the placebo normalization described above). Despite
the fact that $D_{1}$ is associated with lower ADE rate than $D_{2}$
in both direct ($T_{1}$) and indirect ($T_{2}$ \emph{vs} $T_{3}$)
comparisons, the uncorrected mean ADE rate is higher for $D_{1}$
($\frac{40+10}{100+100}=25\%$) than for $D_{2}$ ($\frac{50+22}{100+200}=24\%$).
Actually, it gives a higher weight to $T_{1}$ for $D_{1}$, and $T_{1}$
is associated with a higher overall rate of ADE (possibly due to the
trial conditions).

We correct the data as follows. We compute the overall indirect /
direct patient ratio $r$. In the previous example, $r=\frac{100+200}{100+100}=1.5$.
Then, for each of the compared group $D$, we weighted the direct
comparisons with a factor $k_{dir}(D)$ and the indirect comparisons
with a factor $k_{ind}(D)$ in order to obtain a ratio equal to $r$.
Let us note $\left|D\right|_{dir}$ and $\left|D\right|_{ind}$ the
number of patients in the direct and indirect comparisons for group
$D$, respectively. We have:

\[
k_{dir}(D)=\min\left(1,\frac{\left|D\right|_{ind}}{\left|D\right|_{dir}\times r}\right)
\]

\[
k_{ind}(D)=\min\left(1,\frac{\left|D\right|_{dir}\times r}{\left|D\right|_{ind}}\right)
\]

As previously, the weight given to a patient cannot be higher than
1. In the previous example, we have $k_{dir}(D_{1})=\frac{100}{100\times1.5}=0.667$,
$k_{ind}(D_{1})=1$, $k_{dir}(D_{2})=1$ and $k_{ind}(D_{2})=\frac{100\times1.5}{200}=0.75$.

~

\subsection{Visual analytics}

Let us consider a 26-dimensional numeric dataset, with 2 dimensions
for each ADE category, one for all ADE in the category and the other
for serious ADE only. In this dataset, each group of similar patients
in the query corresponds to a data point. We chose glyphs for the
visualization of the dataset, because they are particularly efficient
for identifying differences between data points. Moreover, dimension
reduction techniques would imply an important information loss (see
section \ref{subsec:Multidimensional-visualization-t}), and parallel
coordinates consider each dimension in the same way, while our 26
dimensions are organized in $13\times2$ corresponding to the 13 ADE
categories and the 2 levels of seriousness.

\begin{figure}
\noindent \begin{centering}
\includegraphics[width=1\columnwidth]{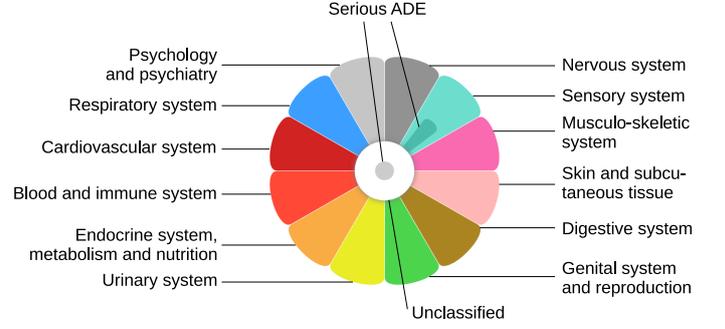}
\par\end{centering}
\caption{\label{fig:fleur}Structure of the flower glyph, with the 13 ADE categories.}
\end{figure}

We adapted flower glyphs for visualizing the per-category and per-seriousness
level rates of ADE observed in a group. Unclassified ADE are represented
by a white circle at the center of the flower, and each of the 12
remnant categories is represented by a colored petal. Both the position
of the petal and its color were chosen in order to facilitate the
understanding and the memorization of the category. For example, nervous
system is at the top and in gray (think of the brain and the ``gray
matter''), while the urinary system is in yellow at the bottom. When
no ``obvious'' colors were available, arbitrary colors were used,
e.g. green for genital system and reproduction. Figure \ref{fig:fleur}
shows the structure of the flower glyph and the color and position
of the 13 ADE categories.

Contrary to what was found in the literature, we encoded the ADE rate
by the area of the petal, and not by its length. The area of the center
circle and the petals is thus proportional to the observed rate of
the corresponding ADE (including both serious and non-serious ADE).
When present, serious ADE were represented by a darker central circle
or a darker smaller petal. The area of this darker region is proportional
to the observed rate of serious ADE in the given category.

This flower glyph takes advantage of the ability of the human vision
to distinguish at least 12 directions, as in an analog clock, and
its higher sensitivity to area rather than to distance \citep{Ware2008_2}.
In addition, compared to a bar chart, the overall triangular shape
of petals associated with area proportionality allows giving more
attention to small values, \emph{i.e.} if the rate of ADE is multiplied
by 2, the length of the petal is multiplied by less than 2, because
the area increases faster than the length. This acts similarly to
a logarithmic scale, although not logarithmic from a mathematical
point of view. Consequently, small values remain visible when much
higher values are present.

We added interactivity to flower glyphs, as follows. When the mouse
is over a region (the central circle or a petal), a popup bubble displays
the ADE category label and the associated rate of ADE and serious
ADE, with the most frequent ADE and serious ADE in the category. When
the mouse is clicked, the webpage is scrolled down to display the
entire list of ADE in this category. Finally, when several glyphs
are present, we added a $\Delta$ (delta) button. When this button
is mouse-hovered, the outline of the selected glyph is drawn as a
wire frame over the other glyphs, in order to facilitate comparison.

\subsection{Implementation details}

The search procedure was implemented with Owlready, which translates
the query into an SQL query. The visual analytics was implemented
in a web platform, using Python 3 with Flask and Owlready, web technology
(HTML, CSS) and Brython, a client-side Python interpreter.

\subsection{Use cases and expert opinions}

Several use cases were designed for the platform. Most of them were
focused on trying to find back already known results, \emph{e.g.}
from meta-analysis, in order to validate our approach. An additional
use case was designed by selecting an indication in the ontology and
comparing the available drugs, in order to search for possible new
insights.

The proposed platform was presented to four experts in drug safety,
from the French drug agency (\emph{Agence Nationnale de Sécurité du
Médicament et des produits de santé}, ANSM), using the use cases.
Then, the comments, opinions and suggestions of the experts were collected
during a focus group session.

\section{\label{sec:Results}Results}

\subsection{The ontology of pain treatment ADE}

\begin{figure}[t]
\noindent \begin{centering}
\includegraphics[width=1\columnwidth]{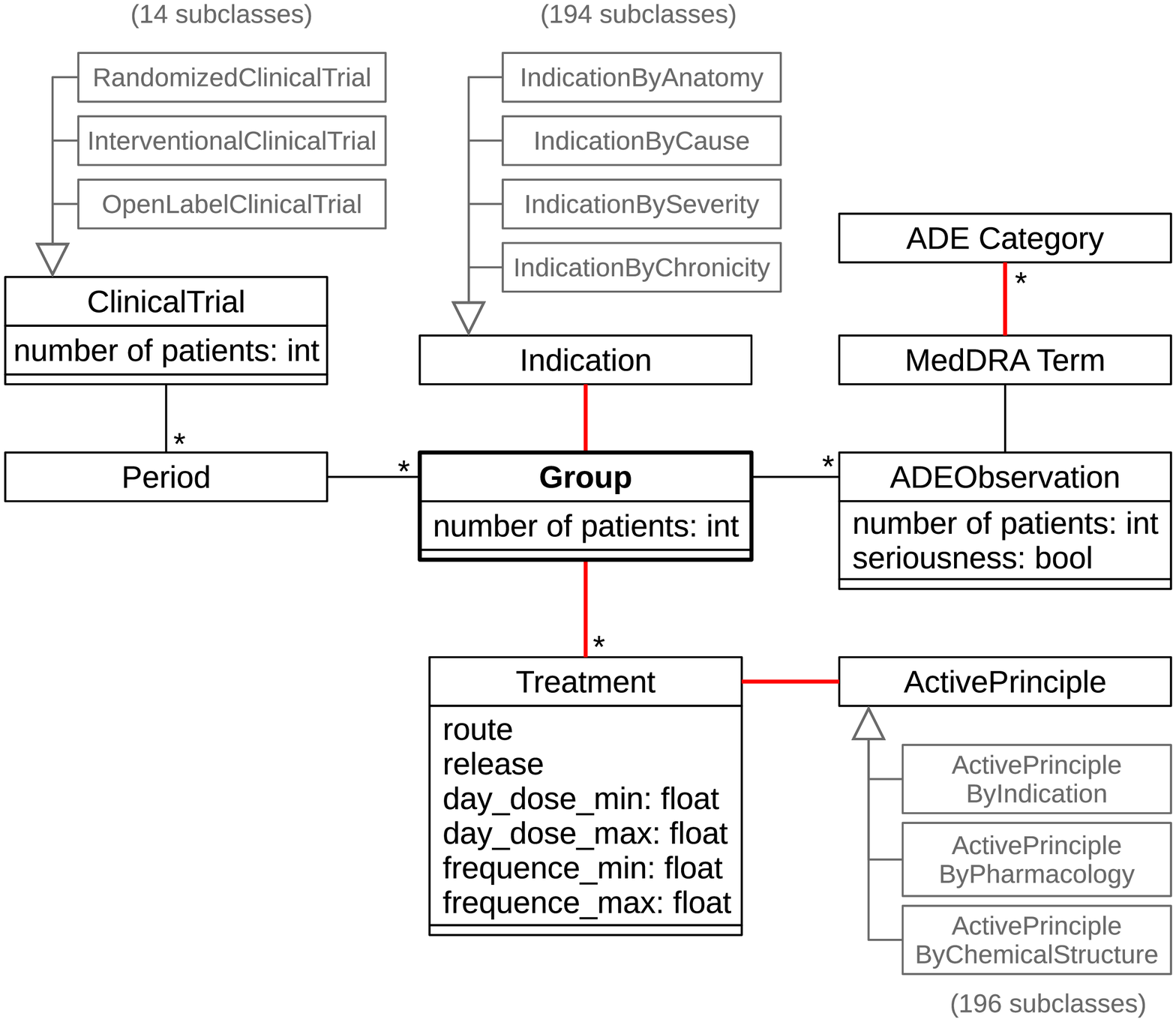}
\par\end{centering}
\caption{\label{fig:onto}General model of the ontology of pain treatment ADE
in UML. Relations in black were extracted automatically, while manual
intervention was required for those in red.}
\end{figure}

Figure \ref{fig:onto} shows the general model of the pain treatment
ADE ontology, in UML (Unified Modeling Language). In this model, Group
is the central class and represents a group of similar patients, in
terms of clinical conditions and treatments received. A clinical trial
contains one or more periods; most trials either include a single
period or an open-label titration period followed by a maintenance
period. A few also include an open-label continuation or pick-up period.
Each period includes one or more comparable group. Each group has
one or more drug treatments, prescribed for one or more indications.
ADE are observed in groups, and are associated with a MedDRA term
and a seriousness Boolean status. 18,090 ADE terms were extracted
from ClinicalTrials.gov. 17,304 (95.7\%) were automatically mapped
to MedDRA terms using their label, the others being associated with
the MedDRA SOC present in ClinicalTrials.gov. ADE with the same MedDRA
term may differ in seriousness, \emph{e.g.} diarrhea might be serious
for newborns but not for adults. Finally, each MedDRA term is associated
with one or two of the 13 ADE categories.

\begin{figure}[!t]
\noindent \begin{centering}
\includegraphics[width=1\columnwidth]{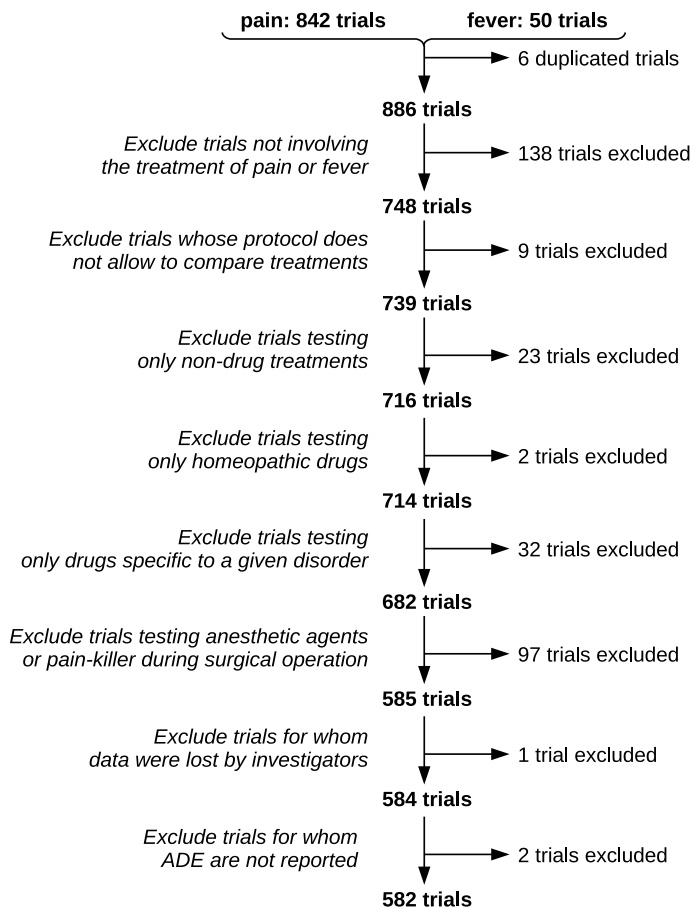}
\par\end{centering}
\caption{\label{fig:ct_flow}Selection of clinical trials for the study.}
\end{figure}

Inheritance is present at three levels: ClinicalTrial, Indication
and ActivePrinciple. Notice that OWL ontologies allow \emph{multiple
instantiation} in addition to multiple inheritance, \emph{i.e.} a
given clinical trial can belong to several classes, \emph{e.g.} RandomizedClinicalTrial
and InterventionalClinicalTrial. In Figure \ref{fig:onto}, relations
in red were extracted from ClinicalTrials.org semi-automatically,
with manual intervention, while those in black were extracted automatically.

\begin{figure}[!t]
\noindent \begin{centering}
\includegraphics[width=0.9\columnwidth]{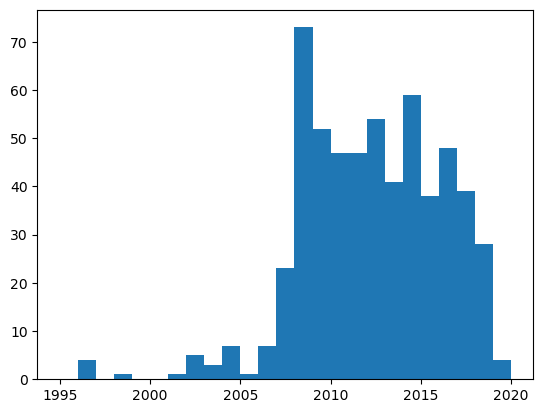}
\par\end{centering}
\caption{\label{fig:histo}Histogram of the completion dates of the selected
trials.}
\end{figure}

Figure \ref{fig:ct_flow} shows the selection and exclusion of clinical
trials during the study. 582 clinical trials were included. In addition
to the criteria mentioned in section \ref{sec:Materials}, one trial
was excluded because results data was not present in ClinicalTrials.gov
(investigators lost data during flooding, NCT01401049), and two trials
because they contain only results related to efficacy. Figure \ref{fig:histo}
shows a histogram of the completion date of the selected trials. Most
trials were completed after 2007.

During the ontology population, 1,394 groups were extracted with 1,653
individual drug treatments (a group may have several treatments, \emph{e.g.}
2 in case of a bitherapy). The Python script extracted the right indication
for 77.3\% of the groups, the right severity for 97.1\%, and the right
chronicity for 71.2\%. It also extracted the right active principle
for 77.7\% of the individual drug treatments, the right release for
90.1\%, the right route for 70.3\%, the right dose for 55.3\%, the
right dose unit for 61.4\%, and the right number of intakes per day
for 74.2\%.

The resulting ontology includes 582 trials, 1,354 groups and 157,665
patients, 201 active principle classes, 194 indication classes, and
148,843 reported individual ADE. It was formalized in OWL 2.0 and
saved in RDF/XML, and it contains 299,341 RDF triples. The ontology
is publicly available at \href{http://www.lesfleursdunormal.fr/static/appliweb/pain/pain_onto.zip}{http://www.lesfleursdunormal.fr/static/appliweb/pain/pain\_onto.zip}.

\begin{figure*}[p]
\noindent \begin{centering}
\includegraphics[width=0.82\textwidth]{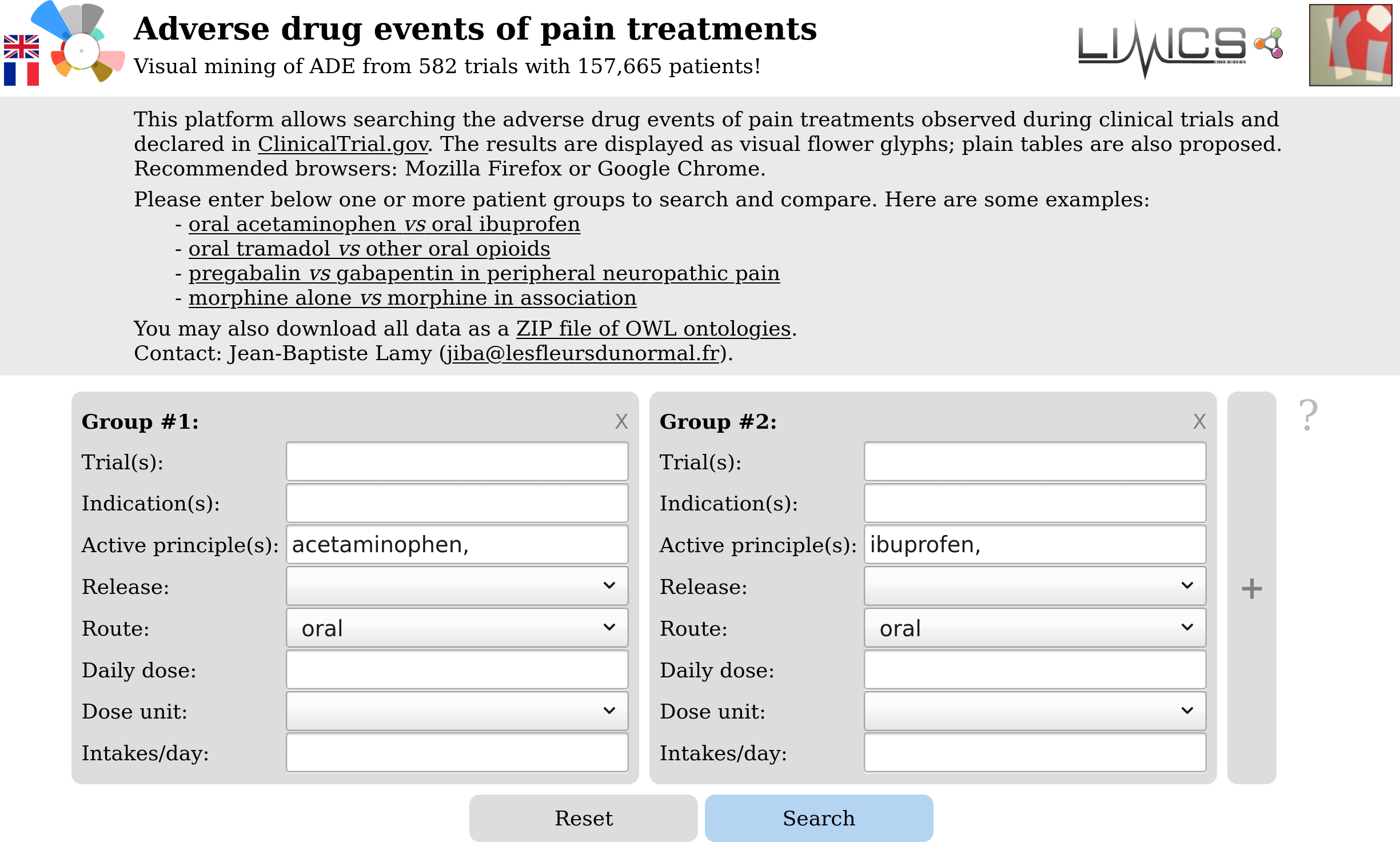}
\par\end{centering}
\caption{\label{fig:search}Screenshot of the search interface.}
\end{figure*}

\begin{figure*}[!p]
\noindent \begin{centering}
\includegraphics[width=0.82\textwidth]{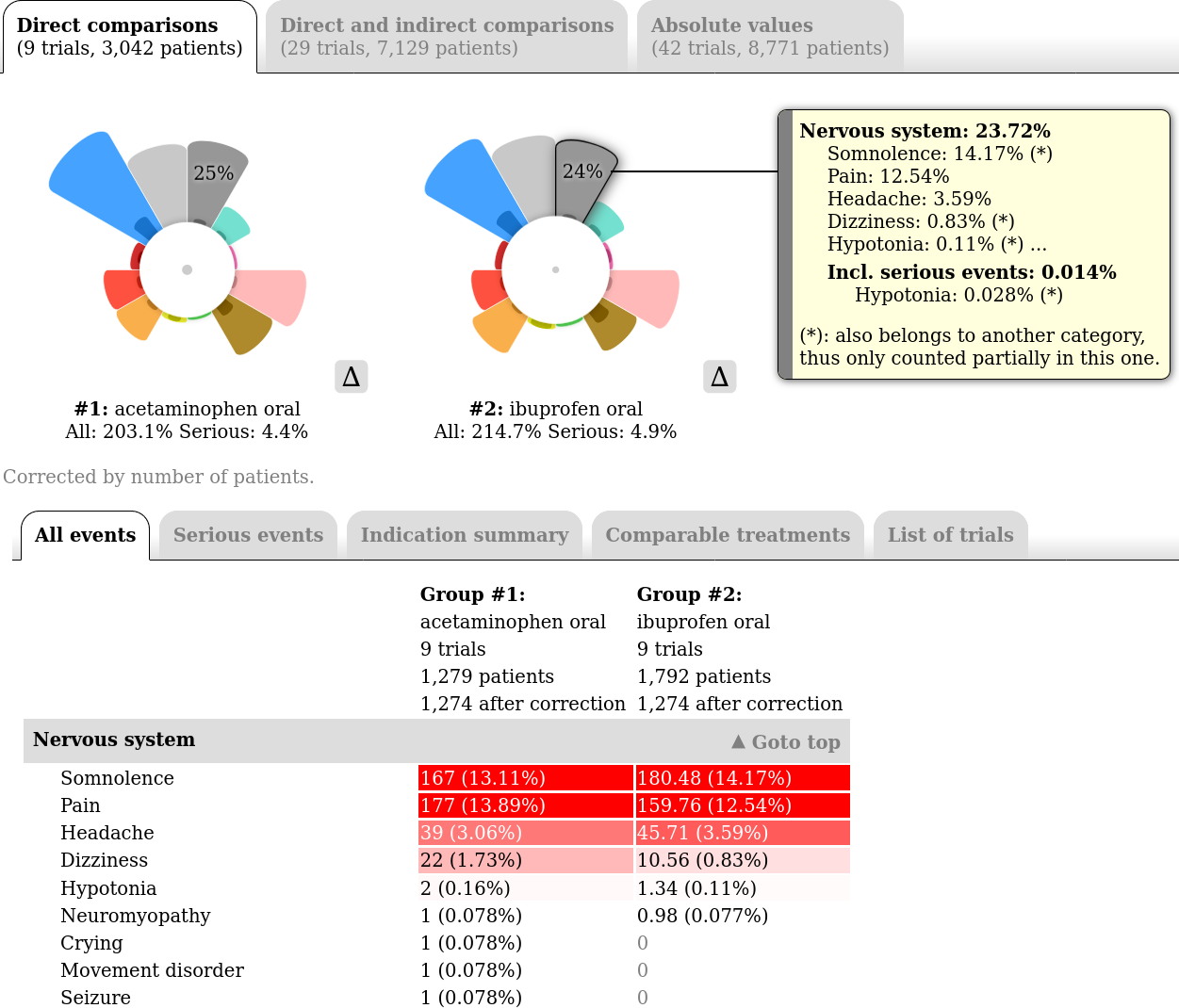}
\par\end{centering}
\caption{\label{fig:results}Screenshot of the results interface, with two
flower glyphs and a bubble showing details on a petal.}
\end{figure*}

\subsection{The data mining web platform}

The proposed platform is available online at \href{http://www.lesfleursdunormal.fr/appliweb/pain}{http://www.lesfleursdunormal.fr/appliweb/pain},
and is fully bilingual (English and French). It has been tested with
both Mozilla Firefox and Google Chrome. Figure \ref{fig:search} shows
a screenshot of the search interface. The user can enter one or more
groups. Each group may contain one or several comma-separated classes
of trial, indication and active principle. Auto-completion is used
for facilitating the entry of trial types, indications and active
principles, and the entire hierarchies can be displayed by clinking
on the field labels. If several active principles are entered for
a given group, the per-drug fields (\emph{i.e.} release, route, daily
dose, dose unit and intakes/day) are subdivided with one field for
each active principle. In the results, each group will be displayed
in a separate flower glyph. In Figure \ref{fig:search}, the user
defined two groups: ``oral acetaminophen'' and ``oral ibuprofen''.

When entering active principles, the ``etc'' special label can be
used for defining an ``open list'' of active principles, \emph{e.g.}
``morphine, etc'' for any treatment that includes morphine (possibly
with other active principles). In addition, when a group's active
principles are more general than a previous group, we automatically
exclude the results of the more specific group from the result of
the more general group. For example, when comparing tapentadol with
opioids, the search will automatically consider ``opioids other than
tapentadol'' (labeled ``other opioid'' in the results, for brevity).

Figure \ref{fig:results} shows a screenshot of the results interface,
after the user performed the query shown in Figure \ref{fig:search}.
The results interface includes three parts, organized vertically.
First, the three tabs at the top of the screen allows selecting the
result sets: direct comparisons, direct and indirect comparisons,
or absolute values. The first one includes only trials with all searched
groups, the second one also includes trials with only some of the
queried groups and a placebo group allowing adjusting the ADE observed
according to the values observed for the placebo (as explained in
section \ref{subsec:Data-correction}), and the third one includes
all groups found independently from the trial they belong to, without
any corrections. The number of trials and patients increase from the
left to the right tab, but the quality of the data decreases. The
``absolute values'' result set is surely not sufficient for drawing
conclusions; however, it is very useful for confirming or invalidating
the conclusions observed in the other result sets, or for obtaining
general trends when the other result sets are not available.

Second, flower glyphs display the rates of the various categories
of ADE, for each group. The user can mouse over the region of the
glyph to obtain more detail in a popup bubble, or click to scroll
down to the entire list of ADE in the chosen category. At the bottom
right of the glyph, the $\Delta$ button allows drawing the outline
of the select glyph on top of the other glyphs. This permits fine
comparison and facilitates the identification of small differences.
At the bottom of the flower glyphs, a line of text summarizes the
corrections that were applied to the data.

Notice the very high ADE rates shown in Figure \ref{fig:results}:
more than 200\%, \emph{i.e.} more than 2 ADE per patients on average.
These rates correspond to the number of events observed during trials,
but may not correspond to the real rate of ADE caused by the drug
in normal clinical use: of course, the ADE rate of acetaminophen and
ibuprofen is not 200\%. In particular, the ADE may be observed during
a long period (several months or even years, \emph{i.e.} for acetaminophen
and ibuprofen, some trials on post-vaccination fever cover all vaccinations
during childhood), and the events observed may be due to the drug
taken, but also to the patient disorders or the conditions of the
trial (\emph{e.g.} ADE may be caused by the vaccine in the previous
example) and to random hazards (\emph{e.g.} an infection may occur
during the trial, unrelated to the treatment). However, the rates
given by the platform are comparable for the ``direct'' and ``direct
+ indirect'' result sets, allowing the comparison of the various
treatment options.

\begin{figure}[!t]
\noindent \begin{centering}
\includegraphics[width=1\columnwidth]{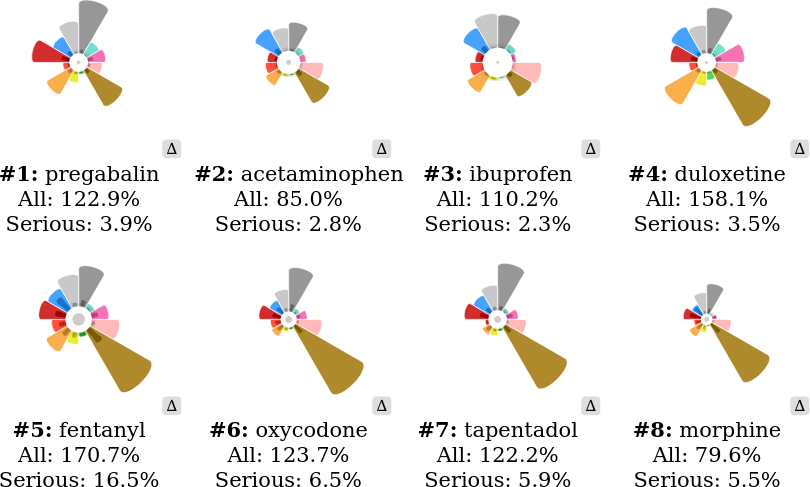}
\par\end{centering}
\caption{\label{fig:glyphs}Flower glyphs corresponding to the 8 most common
pain treatments in the ontology, using the \textquotedblleft absolute
values\textquotedblright{} result set. Hence the glyphs are not comparable
since the patient conditions may differ, but these glyphs give an
idea of the type of ADE one may expect which each drug when used in
its own typical conditions.}
\end{figure}

The size of the flower glyphs reduces when the number of glyphs increases.
Figure \ref{fig:glyphs} shows a ``bouquet'' of flower glyphs for
the 8 most common pain treatments in the ontology. We can observe
that digestive ADE are the most frequent, followed by neurologic ADE.

\begin{figure}[!t]
\noindent \begin{centering}
\includegraphics[width=1\columnwidth]{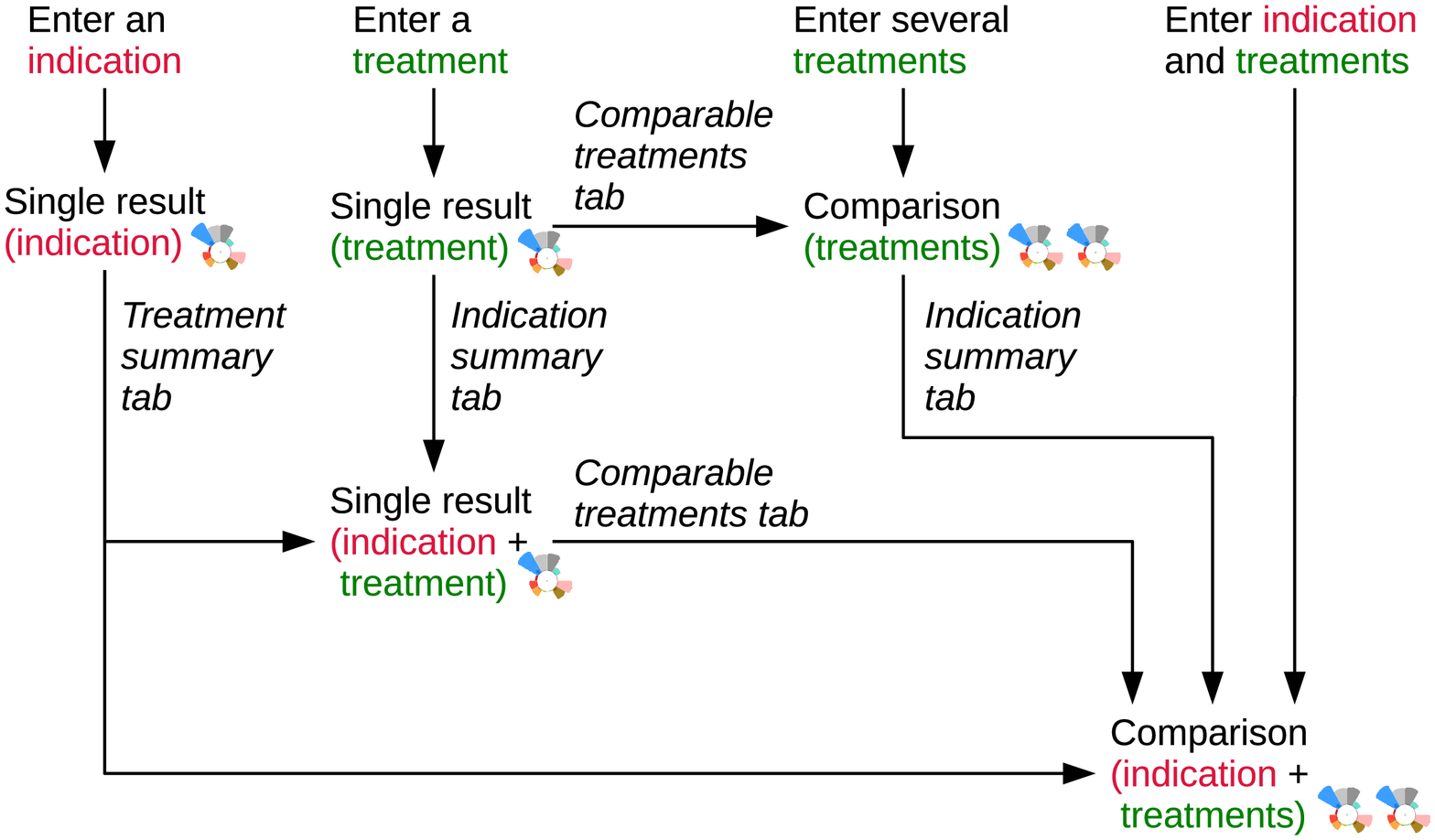}
\par\end{centering}
\caption{\label{fig:trajectoires}Example of typical search strategies in the
platform.}
\end{figure}

Third, a second set of tabs allows displaying various tabular information.
The tabs are: (1) ``All events'': this tab displays the entire list
of all ADE, sorted by category and then by rate. Background reddish
colors are used to represent frequencies visually, from white (0\%)
to red (5\% or more) on a logarithmic scale. (2) ``Serious events'':
this tab is similar to the previous one, but displays only serious
ADE. (3) ``Indication summary'': this tab displays the indications
of the selected trials, ordered by frequency. Radio buttons allow
selecting one of the indications, in order to restrict the search
to the chosen indication. (4) ``Treatment summary'': this tab displays
the list of treatments tested in the selected trials, ordered by frequency.
It is present only if no active principles were entered in the query,
\emph{e.g.} the user may search for ``cancer pain'' and will find
in this tab the most frequently tested treatments for cancer pain.
Checkboxes allow selecting all or part of these treatments, with a
button for performing a new search comparing the selected treatments
in the previously entered indication. (5) ``Comparable treatments'':
this tab displays similar treatments used as comparators in trials,
ordered by frequency. It is present only if active principles were
entered in the query. For example, the user may search for ``tramadol'',
and will get in this tab the list of treatments compared to tramadol.
Checkboxes allow selecting some comparable treatments, with a button
for adding the selected treatment into the current search. (6) ``List
of trials'': this tab lists all the trials in the current result
set, with links to pages displaying the ADE observed in each trial.
It also displays the per-trial rate of ADE for each group, allowing
the identification of outlier trials. Finally, it includes checkboxes,
allowing the exclusion of some trials from the result set. Figure
\ref{fig:trajectoires} shows several search strategies made possible
using tabs 3-5.

\subsection{Use cases}

\subsubsection{Acetaminophen \emph{vs} ibuprofen}

Acetaminophen, ibuprofen and aspirin are the three main OTC (Over
The Counter, \emph{i.e.} without medical prescriptions) painkillers
taken by patients. However, aspirin is no longer recommended for such
use. Thus, we found no recent trial involving aspirin as a painkiller.
Figure \ref{fig:results} shows the direct comparison of oral acetaminophen
\emph{vs} oral ibuprofen, including 9 trials and 3,042 patients\footnote{Available online at \href{http://www.lesfleursdunormal.fr/appliweb/pain?group_1_ap=acetaminophen&group_1_route=oral&group_2_ap=ibuprofen&group_2_route=oral}{http://www.lesfleursdunormal.fr/appliweb/pain? group\_1\_ap=acetaminophen\&group\_1\_route=oral\&group\_2\_ap=ibuprofen \&group\_2\_route=oral}}.
The results show slightly fewer ADE with acetaminophen (203.1\% \emph{vs}
214.7\%), as well as for serious ADE (4.4\% \emph{vs} 4.9\%). In particular,
acetaminophen is associated with fewer unclassified ADE (mostly fevers,
54\% \emph{vs} 70\%). Nevertheless, the difference is low and probably
not significant, and the ADE profile of both drugs is very similar.
These results are consistent with a meta-analysis that concluded,
10 years ago, that acetaminophen and ibuprofen are equally safe \citep{Pierce2010},
but the results presented here include more recent trials.

Two main indications are present in the ``Indication summary'' tab:
post-vaccination fever in children and pain following dental extraction.
Interestingly, separate analyses (easily available using the radio
buttons in the tab) suggest that acetaminophen has fewer ADE in the
former indication and ibuprofen in the latter.

\begin{figure}[!t]
\noindent \begin{centering}
\includegraphics[width=1\columnwidth]{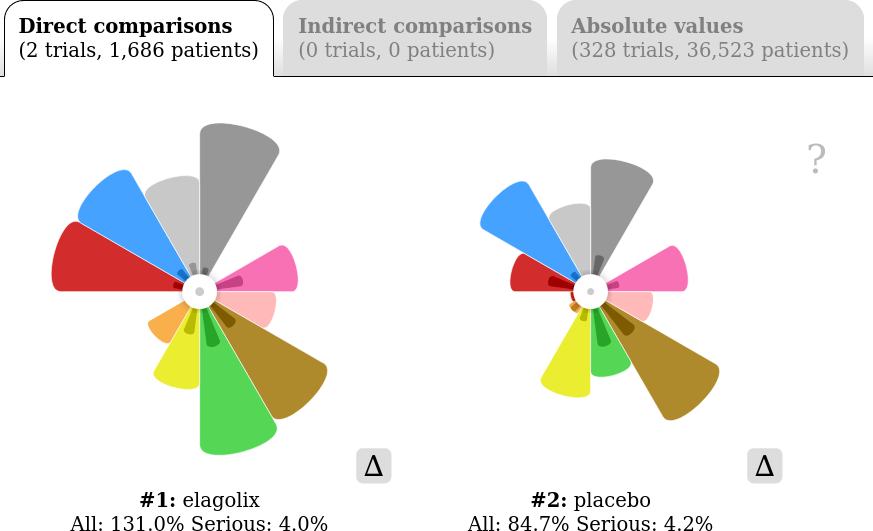}
\par\end{centering}
\caption{\label{fig:elagolix}Flower glyphs comparing elagolix to placebo.}
\end{figure}

\subsubsection{Elagolix}

Elagolix is a gonadotropin-releasing hormone antagonist drug approved
in 2018 in US, for the treatment of pain associated with endometriosis
in women \citep{Lamb2018}. However, elagolix is known to be associated
with frequent ADE such as hot flushes \citep{Vercellini2019}.

After searching for elagolix alone, the ``comparable treatments''
tab shows that elagolix was only compared with placebo, during two
trials. The tab's checkbox permits the addition of placebo to the
query, leading to Figure \ref{fig:elagolix}\footnote{Available online at \href{http://www.lesfleursdunormal.fr/appliweb/pain?group_1_ap=elagolix&group_2_ap=placebo}{http://www.lesfleursdunormal.fr/appliweb/pain? group\_1\_ap=elagolix\&group\_2\_ap=placebo}}.
We can clearly see that the ADE rate is higher for elagolix, especially
for cardiovascular and genital/reproductive ADE (corresponding to
the red and green petals). In particular, elagolix is associated with
34\% risk of hot flush (classified in both cardiovascular and genital/reproductive
ADE categories, and thus counting for half in each) and a 5.6\% risk
of amenorrhea. For placebo, these risks are 8.6\% and 0.27\%, respectively.
These values can be easily obtained in the petal popup bubbles. These
results clearly illustrate the high ADE rate with elagolix.

\begin{figure}[!t]
\noindent \begin{centering}
\includegraphics[width=1\columnwidth]{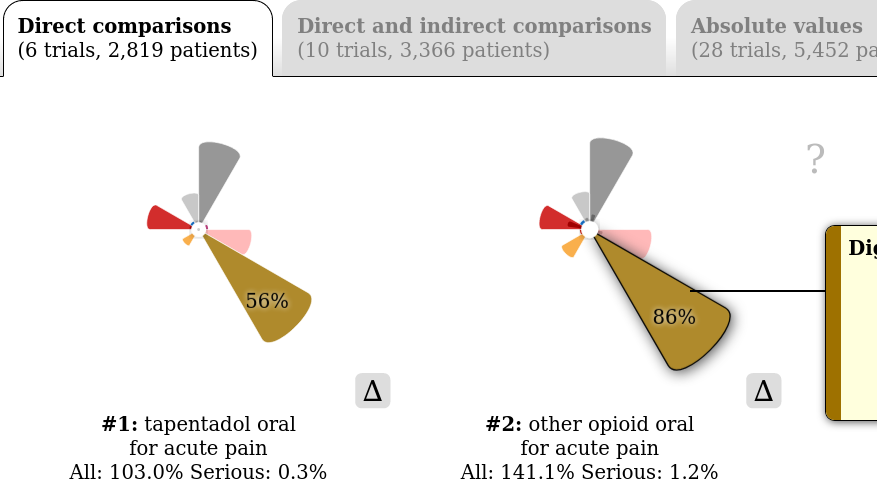}
\par\end{centering}
\caption{\label{fig:tapentadol}Flower glyphs comparing oral tapentadol to
other oral opioids for acute pain.}
\end{figure}

\subsubsection{Tapentadol}

As said in the introduction, a recent meta-analysis showed that tapentadol
was associated with less digestive ADE than other opioids when prescribed
for acute pain \citep{Wang2020}. Figure \ref{fig:tapentadol} shows
the direct comparison of oral tapentadol with other oral opioids\footnote{Available online at \href{http://www.lesfleursdunormal.fr/appliweb/pain?group_1_indication=acute\%20pain&group_1_ap=tapentadol&group_1_route=oral&group_2_indication=acute\%20pain&group_2_ap=opioid&group_2_route=oral}{http://www.lesfleursdunormal.fr/appliweb/pain? group\_1\_indication=acute pain\&group\_1\_ap=tapentadol\&group\_1\_route=oral \&group\_2\_indication=acute pain\&group\_2\_ap=opioid\&group\_2\_route=oral}}.
Six trials were found, involving 2,819 patients. On the flower glyphs,
we can clearly see the difference in digestive ADE (56\% \emph{vs}
86\%).

Contrary to the meta-analysis, the other opioids involved in the comparison
are not limited to a single option, but include both oxycodone and
morphine, as shown in the ``Treatment summary'' tab. One may perform
additional searches to compare tapentadol to oxycodone and morphine
separately, showing that tapentadol has fewer digestive ADE in each
comparison.

\begin{figure}[!t]
\noindent \begin{centering}
\includegraphics[width=1\columnwidth]{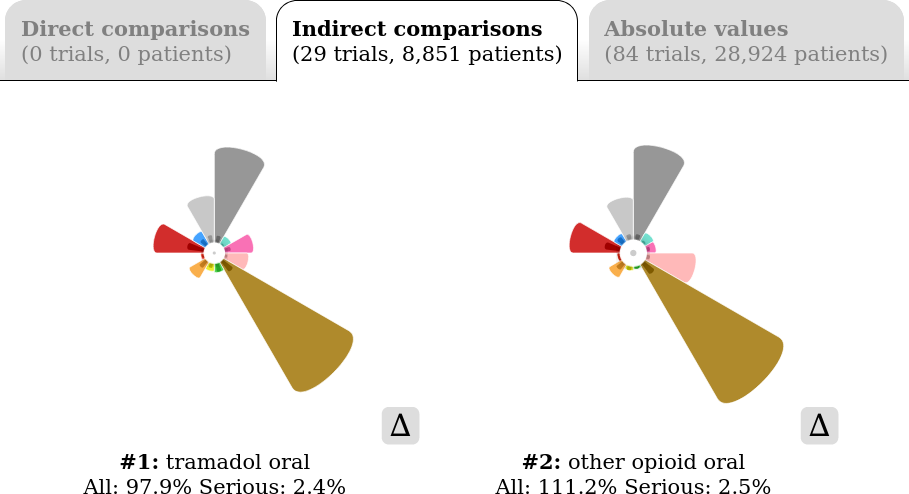}
\par\end{centering}
\caption{\label{fig:tramadol}Flower glyphs comparing tramadol to other opioids.}
\end{figure}

\subsubsection{\label{subsec:Tramadol}Tramadol}

Tramadol is an opioid painkiller used to treat moderate pain. It is
classified as a level-2 painkiller (out of 3 levels). Tramadol was
considered as having fewer risks of ADE compared with other more powerful
opioids. However, it appeared that tramadol was involved in an important
number of ADE, including serotoninergic syndrome \citep{Beakley2015},
vomiting and sleep disorders, and was at risks of misuses and dependence.
A recent study involving Egyptian students showed that 12.3\% of them
were taking tramadol, and that 30\% of the tramadol-consumers had
dependence \citep{Bassiony2018}. Another study showed that patients
receiving tramadol after surgery had similar, or even higher, risks
of prolonged opioid use compared with patients receiving other opioids
\citep{Thiels2019}. In France, the French drug agency (\emph{Agence
Nationale de Sécurité du Médicament et des produits de santé}, ANSM)
recently published a communicate\footnote{``TRAMADOL : une mesure pour limiter le mésusage en France - Point
d'information'', 16/01/2020, \href{https://www.ansm.sante.fr/S-informer/Points-d-information-Points-d-information/TRAMADOL-une-mesure-pour-limiter-le-mesusage-en-France-Point-d-information}{https://www.ansm.sante.fr/S-informer/Points-d-information-Points-d-information/TRAMADOL-une-mesure-pour-limiter-le-mesusage-en-France-Point-d-information}}, and reduced the maximum prescription duration for tramadol, from
12 to 3 months.

Figure \ref{fig:tramadol} shows the comparison of oral tramadol with
other oral opioids\footnote{Available online at \href{http://www.lesfleursdunormal.fr/appliweb/pain/?group_1_ap=tramadol&group_1_route=oral&group_2_ap=opioid&group_2_route=oral}{http://www.lesfleursdunormal.fr/appliweb/pain/? group\_1\_ap=tramadol\&group\_1\_route=oral\&group\_2\_ap=opioid\&group\_2\_route=oral}}.
Here, no direct comparisons were found; this was expected since tramadol
is not considered as comparable with level-3 opioids like morphine.
The indirect comparison result set, normalized by placebo, includes
29 trials (5 for tramadol and 24 for other opioids) and 8,851 patients.
It clearly appears that the ADE profile of tramadol is very similar
to the one of other oral opioids, with a high ADE rate and many digestive
and nervous ADE. This supports the facts that tramadol is potentially
as dangerous as other opioids.

On the contrary, when looking at the details of the psychological
ADE, no occurrences of drug abuse were found for tramadol. But this
may be due to the conditions of clinical trials, which often have
relatively short durations, and very controlled doses and protocols
that limit the risk of drug abuse. Moreover, certain trials may exclude
participants with a history of substance abuse.

\begin{figure}[!t]
\noindent \begin{centering}
\includegraphics[width=1\columnwidth]{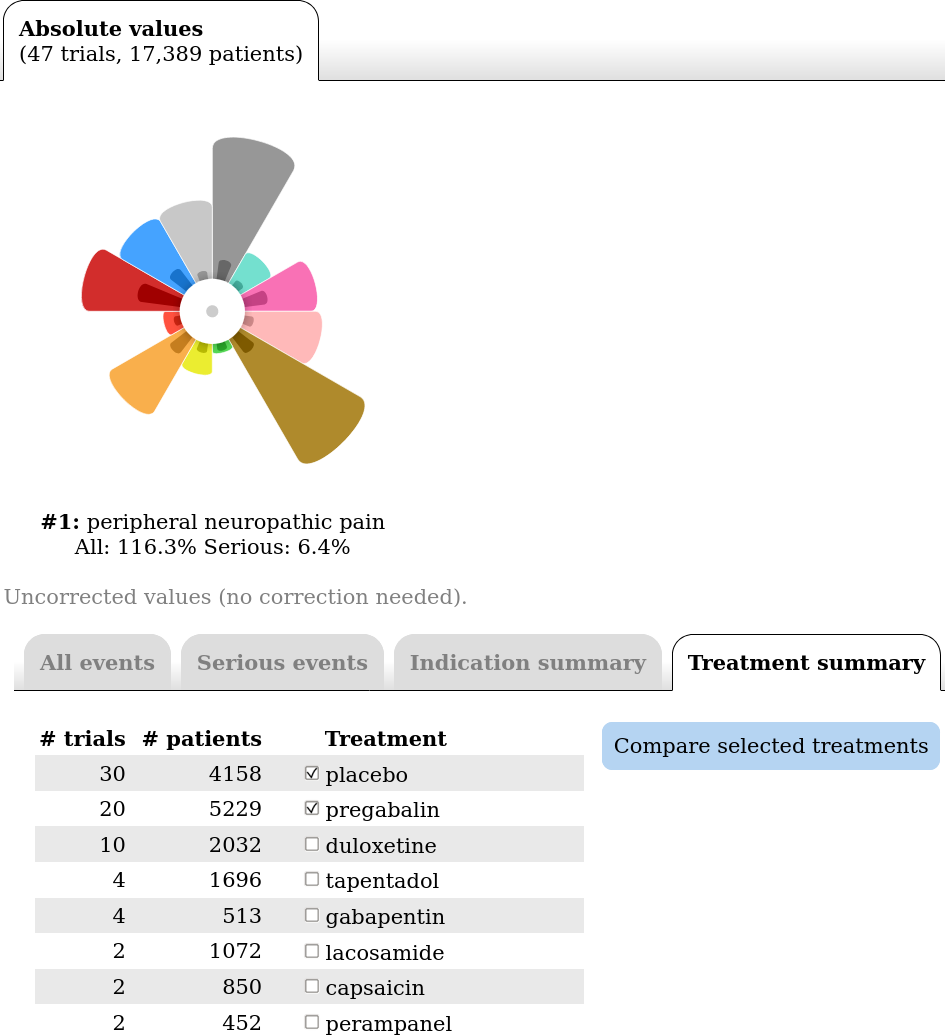}
\par\end{centering}
\caption{\label{fig:douleur_neuro_peri}ADE observed during the treatment of
peripheral neuropathic pain, and the treatment summary.}
\end{figure}

\subsubsection{Peripheral neuropathic pain and gabapentinoids}

Peripheral neuropathic pain is a pain caused by damages to peripheral
nerves. Typical causes of damages are diabetic neuropathy, postherpetic
neuralgia and trauma. The 2010 guidelines on the pharmacological treatment
of neuropathic pain \citep{Attal2010} from EFNS (European Federation
of Neurological Societies) recommends duloxetine, pregabalin, gabapentin,
tricyclic antidepressants or venlafaxine as first-line treatment for
painful polyneuropathy (including diabetic neuropathic pain) and pregabalin,
gabapentin, tricyclic antidepressants or lidocaine for post-herpetic
neuralgia. In 2015, a systematic review and meta-analysis recommended
gabapentin, gabapentin enacarbil, pregabalin, duloxetine, venlafaxine
or tricyclic antidepressants as first-line treatment for neuropathic
pain in adults \citep{Finnerup2015}. In 2013, a meta-analysis concluded
that gabapentin has the most favorable balance between efficacy and
safety for diabetic neuropathic pain \citep{Rudroju2013}, but since
this study did not analyze efficacy and safety separately, it is difficult
to know whether this conclusion is due to efficacy, safety or a mix
of both.

Figure \ref{fig:douleur_neuro_peri} shows the results obtained when
searching for ``peripheral neuropathic pain'' in the platform\footnote{Available online at \href{http://www.lesfleursdunormal.fr/appliweb/pain?group_1_indication=peripheral\%20neuropathic\%20pain}{http://www.lesfleursdunormal.fr/appliweb/pain? group\_1\_indication=peripheral neuropathic pain}}.
It shows the ADE caused by all drug treatments prescribed for peripheral
neuropathic pain in the ontology. The ``Treatment summary'' tab
can be used for finding the most prescribed active principles in that
indication.

\begin{figure}[!t]
\noindent \begin{centering}
\includegraphics[width=1\columnwidth]{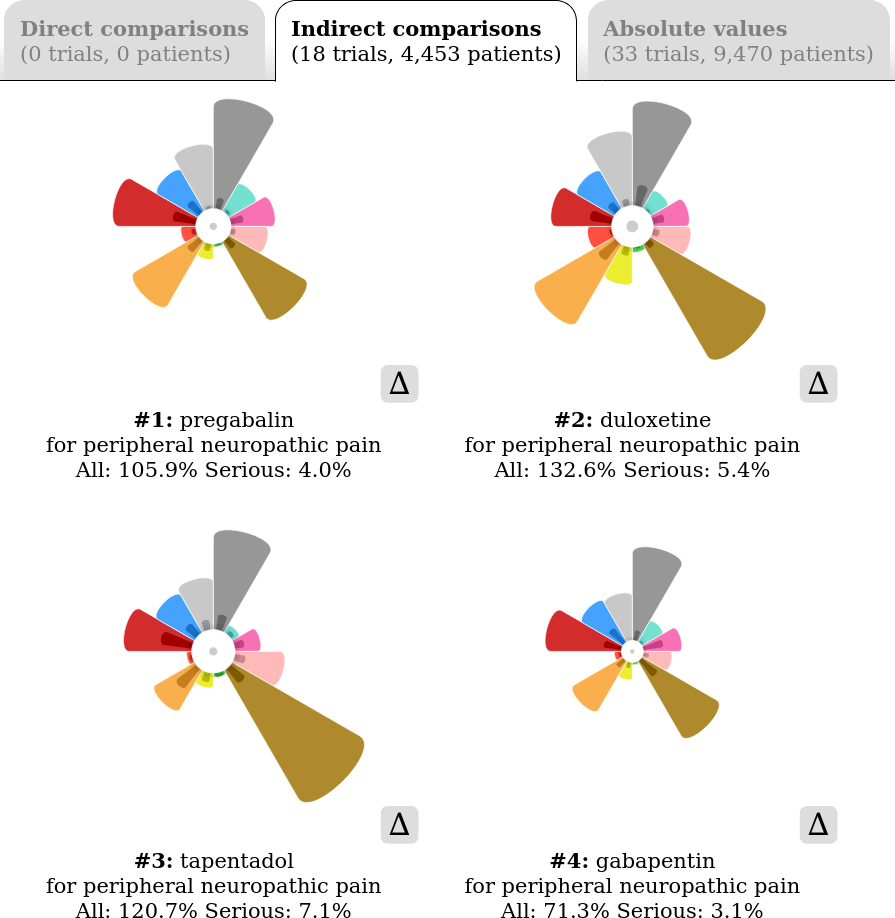}
\par\end{centering}
\caption{\label{fig:douleur_neuro_peri2}Indirect comparison of the four main
treatments for peripheral neuropathic pain.}
\end{figure}

Figure \ref{fig:douleur_neuro_peri2} shows the indirect comparison
obtained after selecting the first four treatments in the ``Treatment
summary'' tab (excluding placebo)\footnote{Available online at \href{http://www.lesfleursdunormal.fr/appliweb/pain?group_1_indication=peripheral\%20neuropathic\%20pain,&group_1_ap=pregabalin&group_2_indication=peripheral\%20neuropathic\%20pain,&group_2_ap=duloxetine&group_3_indication=peripheral\%20neuropathic\%20pain,&group_3_ap=tapentadol&group_4_indication=peripheral\%20neuropathic\%20pain,&group_4_ap=gabapentin}{http://www.lesfleursdunormal.fr/appliweb/pain? group\_1\_indication=peripheral neuropathic pain,\&group\_1\_ap=pregabalin \&group\_2\_indication=peripheral neuropathic pain,\&group\_2\_ap=duloxetine \&group\_3\_indication=peripheral neuropathic pain,\&group\_3\_ap=tapentadol \&group\_4\_indication=peripheral neuropathic pain,\&group\_4\_ap=gabapentin}}.
There are 12 trials for pregabalin, 3 for duloxetine, 2 for tapentadol
and 1 for gabapentin (Notice that these numbers are lower than those
shown in Figure \ref{fig:douleur_neuro_peri}, because only trials
with a placebo could be included in the indirect comparison of Figure
\ref{fig:douleur_neuro_peri2}). We can see that all treatments are
not equal in terms of ADE, despite the fact that three of them were
recommended as first-line treatment (pregabalin, duloxetine and gabapentin).
Duloxetine and tapentadol have the highest ADE rates, with a lot of
digestive ADE for both, and of endocrine / metabolic / nutritional
ADE for duloxetine, in addition to the nervous ADE present with all
four treatments.

\begin{figure}[!t]
\noindent \begin{centering}
\includegraphics[width=1\columnwidth]{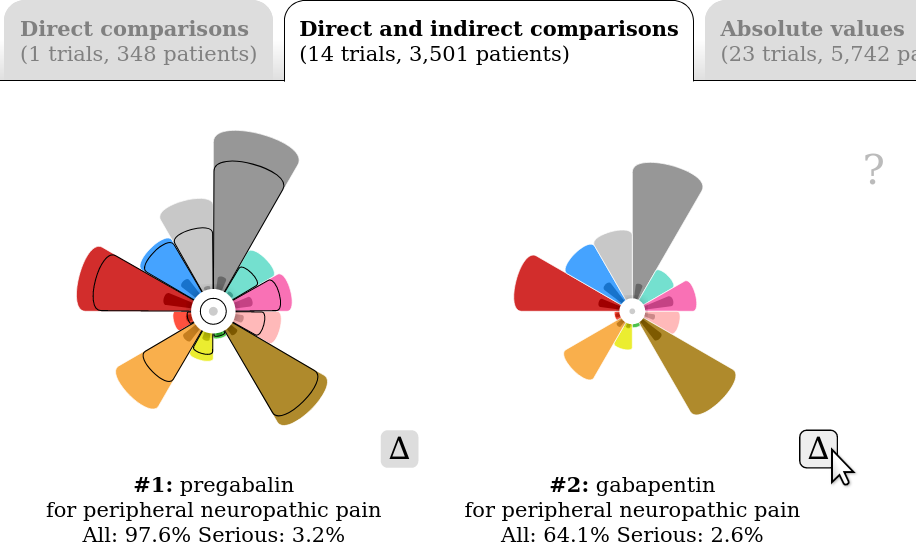}
\par\end{centering}
\caption{\label{fig:pregabaline}Indirect comparison of pregabalin and gabapentin.
We activated the $\Delta$ button of gabapentin, thus showing the
outline of the gabapentin glyph on top of the other one.}
\end{figure}

Pregabalin and gabapentin seems better tolerated. Both belong to gabapentinoids,
a family of antiepileptic drugs which are also prescribed for pain.
Their mechanism of action is not fully known, but an auxiliary subunit
of voltage-gated calcium channels seems to be involved \citep{Taylor2020}.
In Figure \ref{fig:douleur_neuro_peri2}, gabapentin has the fewest
ADE, but the evidence is weak because the gabapentin group includes
only 1 trial and 221 patients. Figure \ref{fig:pregabaline} shows
a direct and indirect mixed comparison of pregabalin \emph{vs} gabapentin,
including 13 trials for pregabalin and 2 for gabapentin (one trial
compare pregabalin to gabapentin with no placebo, and was thus not
present in the previous indirect comparison)\footnote{Available online at \href{http://www.lesfleursdunormal.fr/appliweb/pain?group_1_indication=peripheral\%20neuropathic\%20pain&group_1_ap=pregabalin&group_2_indication=peripheral\%20neuropathic\%20pain&group_2_ap=gabapentin&tab=1}{http://www.lesfleursdunormal.fr/appliweb/pain? group\_1\_indication=peripheral neuropathic pain\&group\_1\_ap=pregabalin \&group\_2\_indication=peripheral neuropathic pain\&group\_2\_ap=gabapentin\&tab=1}}.
The flower glyphs show that both drugs have a similar ADE profile,
but with fewer ADE for gabapentin (64.1\% \emph{vs} 97.6\%). This
suggests that gabapentin might be preferable in terms of safety. It
also suggests that performing a comparative trial between pregabalin
and gabapentin, or a network meta-analysis, may be a valuable study.

\subsection{Expert opinions}

The four experts were very interested by this work. They found the
proposed platform innovative and original, both in its approach and
its presentation. They agreed that trial data are insufficiently exploited
today by computer programs, and they found the visual interface easy
to understand and very nice.

They made several interesting suggestions. First, one expert asked
for additional statistical computations, such as relative risks. Second,
they suggested the use of the platform for approved drugs, but also
before drug approval, in order to help regulators to take the decision
to approve (or not) a new drug by comparing its adverse event profile
with the other drugs already available in the same indication. Third,
they also proposed to compare the adverse event profile of a given
drug in various indications, when a drug is initially approved for
an indication and then prescribed in others. For example, it might
be interesting to compare the ADE of gabapentinoids when prescribed
for epilepsy and for pain: since they were initially approved for
epilepsy, their summaries of product characteristics (SPC) describe
the ADE observed in epilepsy studies, but they might differ from those
occurring when prescribed for pain. Finally, they suggested using
the flower glyph to visualize the adverse effects described in SPC,
and to compare them with those observed during clinical trials.

\section{\label{sec:Discussion}Discussion}

In this paper, we described how ADE mentioned in publicly available
clinical trial results could be integrated in a data mining platform
for drug safety. This platform relies on semantic web technologies
for allowing queries at various levels of granularity, and on visual
analytics, and in particular flower glyphs, for the presentation of
the results. We illustrated the interest of this approach through
case studies, and we showed that we were able to find back results
already known in the literature (\emph{e.g.} regarding elagolix, tapentadol
and tramadol) but also to suggest new results (\emph{e.g.} on gabapentinoids)
that would require further confirmation. The key points of this work
are the original approach producing fact-based evidence on drug safety,
by applying data science to trial data, and the association of semantic
methods with visual analytics.

We found that there is a strong synergy between the semantic and the
visual nature of the proposed data mining platform. The semantic nature
allows searching the entire dataset in many ways, using the links
and options proposed in the various tabs. The visual nature makes
the platform ludic and enjoyable, thus giving the user the desire
to test the many possibilities offered by the platform.

A potential problem in meta-analyses is the \emph{confirmation bias}
\citep{Goodyear-Smith2012}, \emph{i.e.} the tendency to search for,
interpret and favor information in a way that confirms the investigator
hypothesis. Since meta-analyses are performed with a hypothesis in
mind, this hypothesis might bias the process. On the contrary, in
the present work, the selection of clinical trials and their coding
were performed in a systematic manner, without aiming at answering
a specific medical question. In addition, the query procedure and
the visualization are fully automatic. As a consequence, this might
reduce the confirmation bias.

The main limitation of the proposed approach is that it highly depends
on the availability and the quality of clinical trial data. In particular,
limited data are available for older drugs. For example, aspirin has
been used for treating pain for long, but is no longer recommended.
As a consequence, there is no recent trial involving aspirin as a
painkiller, and thus we have no data in the proposed system. Thus
our platform fails to identify the risks associated with the use of
aspirin as a painkiller. Similarly, we have seen in section \ref{subsec:Tramadol}
that limited data was available for situations of drug abuse or drug
overdose.

Another limitation is the training needed for using the proposed platform
properly. The platform was presented to the experts during a guided
tour. It includes contextual helps, but the ability of a naive user
to understand it without the guided tour still need to be evaluated.

The performances of the automatic scripts for populating the ontology
from the trial registry are not good enough to allow a fully automatic
process. Consequently, an automatic update of the ontology is not
possible. Most sophisticated Natural Language Processing (NLP) methods,
\emph{e.g.} Recurrent Neural Network (RNN) \citep{Mikolov2010}, might
achieve better performances. These machine learning methods require
a manually labeled training dataset, which was not available in our
case. However, the ontology populated here could now be used as a
training set, and, in future work, we plan to experiment RNN based
NLP techniques.

The model we proposed for structuring trials matches the needs of
pain treatments, but remains very limited. A more complex model would
be required for generalizing the platform beyond pain treatment, \emph{e.g.}
a model able to represent chemotherapy regimen or time of drug intake.
We designed specific classifications for indications and active principles,
instead of reusing existing terminologies such as ATC, ICD10 or SNOMED
CT. This was motivated by the known limits of publicly available classifications
(\emph{e.g.} ATC and ICD10 do not have multiple inheritance and thus
would be more limited for searching) and the fact that more sophisticated
terminologies are not publicly available (\emph{e.g.} SNOMED CT in
our country, France). Nevertheless, in the near future, we plan to
map those specific classifications to existing classifications, such
as ATC and ICD10.

In the literature, the closest use of flower glyphs is the work of
Pilato \emph{et al.} for the analysis of social sensing on Twitter
\citep{Pilato2016}. The authors proposed flower glyphs with 7 colored
petals and a center part, corresponding to various emotions (joy,
fear,...). In the present work, we extended flower glyphs for the
visualization of all \emph{vs} serious ADE, using an inner, darker,
petal. Color blind people may not be able to distinguish the various
colors of flower glyphs. However, the information carried by the color
is redundant with the information carried by the orientation of the
petal, thus color blind people should still be able to use flower
glyphs efficiently. Most colors on the flower glyph are culturally
independent (\emph{e.g.} red for blood). A notable exception is the
color of the ``skin and subcutaneous tissue'' category.

The proposed platform is limited to descriptive and visual analysis,
but does not perform statistical tests, \emph{e.g.} in order to test
whether a difference observed in ADE rate is significant or not. While
theoretically feasible, implementing statistical tests raises a problem:
each test is associated with a risk $\alpha$ (usually 5\%) and a
risk $\beta$, and multiplying the test cumulates these risks. It
is commonly accepted that, above 5-10 tests, a correction is necessary.
Therefore, we decided not to propose statistical tests for now. For
indirect comparison, we used placebo as a reference. This works well
for mild to moderate pain, however, for severe pain such as cancer
pain, placebo may not be a valid option from an ethical point of view.
In this case, other references should be considered, \emph{e.g.} morphine.

\section{\label{sec:Conclusion}Conclusion}

In conclusion, we proposed a method and a platform for the analysis
of adverse events observed during clinical trials, and published in
trial registries. We applied the platform to pain treatment, and we
showed that we were able to obtain results already known from meta-analyses,
but also to suggest new insights. These results interested drug safety
experts.

This work opens many perspectives for future research. First, the
automatization of the ontology population from trial registries could
be improved, \emph{e.g.} using deep learning and text mining for extracting
drug treatments and their indication. Second, the proposed system
could be extended to the visualization of patient outcomes in trials,
in order to evaluate drug treatment efficacy, or to other data sources,
such as ADE declared in pharmacovigilance databases, or ADE observed
in real-world prescription data such as health records and OHDSI (Observational
Health Data Sciences and Informatics). In this context, the proposed
system should be connected to a data warehouse, in addition to an
ontology. Third, flower glyphs could be adapted to the presentation
of the rate of potential adverse effects described in drug summaries
of product characteristics (SPC) or patient leaflets. Fourth, the
proposed approach could be applied to other medical domains beyond
pain treatment. Finally, the use of the proposed web platform could
be experimented in medical initial and continuing education, or associated
with decision support tools for prescriptions.

\bibliographystyle{elsarticle-num}
\bibliography{biblio_ama}

\end{document}